\documentclass{article}
\usepackage{graphicx} 
\usepackage{tocbibind}
\usepackage{amsmath,amssymb,amsthm}
\usepackage{amsthm}
\usepackage[style=ieee]{biblatex}
\usepackage{tikz-cd}
\usetikzlibrary{arrows.meta, decorations.pathmorphing}
\newtheorem{theorem}{Theorem}[section]
\newtheorem{definition}{Definition}[section]
\newtheorem{corollary}{Corollary}[theorem]

\theoremstyle{definition}
\theoremstyle{remark}

\title{Braid Group Representations and Defect Operators in AdS/CFT Correspondence}
\author{Tzu-Miao Chou }
\date{April 2025}

\begin{document}

\maketitle
\begin{abstract}

This paper investigates the connection between braid group representations, defect operators, and holography within the AdS/CFT framework. It focuses on the correspondence between bulk Wilson loops and boundary defect operators, emphasizing how braid group representations map to these operators. The study also explores fusion and braiding operations in modular tensor categories, which are crucial for understanding anyons in topological quantum field theories. By providing a unified framework, this work bridges the gap between bulk and boundary physics and offers insights into the holographic realization of topological defects. The results suggest new avenues for research in holographic anyons and their applications in quantum field theory and condensed matter physics.

\end{abstract}

\newpage
\tableofcontents
\newpage
\section{Introduction}
\subsection{Motivation}
The study of topological phases of matter has illuminated the profound role that nontrivial braiding statistics can play in quantum systems. Anyons, as particle-like excitations characterized by representations of braid groups, exemplify how topological information becomes dynamically relevant even in low-dimensional field theories. Beyond condensed matter systems, the theoretical framework surrounding anyons has also found deep connections to quantum gravity and gauge theories, notably through Chern-Simons topological quantum field theories.\cite{Witten1998}

In the context of the AdS/CFT correspondence, three-dimensional gravity with a negative cosmological constant admits a Chern-Simons formulation, suggesting that topological degrees of freedom might similarly govern aspects of the bulk geometry. Moreover, the interplay between bulk topological objects and boundary operators hints at a rich structure of holographic dualities, where non-local features in the bulk could be captured by localized, yet nontrivial, boundary observables.\cite{GaiottoKapustin2009}

Despite significant advances, the precise mechanism through which bulk braiding structures manifest themselves on the boundary remains elusive. While Wilson loops and defect operators have individually been studied within AdS/CFT\cite{Aharony2000}, a systematic understanding of their mutual correspondence, particularly through the lens of braid group representations, has yet to be fully developed. This work aims to uncover how braid group structures intrinsic to the bulk can be precisely mapped into the operator algebra of the boundary conformal field theory.

\subsection{Challenges in Understanding Bulk Braid Structures and Boundary Defect Operators}
One of the central challenges in establishing a bulk-boundary dictionary for braid group structures lies in the non-local nature of braiding processes. In contrast to local bulk excitations, braiding involves global topological data, which must somehow be translated into an algebraic or geometric structure on the boundary. Identifying which class of boundary operators captures this non-local information—and how such operators transform under bulk braiding—is a subtle question that demands a careful analysis.

Furthermore, defect operators in conformal field theory, such as twist operators or codimension-two defects, exhibit a rich variety of behaviors, depending on their embedding and fusion properties. It is not a priori clear which types of defects, if any, can precisely record the topological features associated with bulk braiding. The lack of a direct, general construction linking bulk Wilson loop configurations to boundary defects presents a major conceptual gap in the understanding of holographic duality in topological sectors.\cite{Polchinski1998}

These difficulties are compounded by the technical subtleties of working in asymptotically AdS spacetimes, where the choice of gauge, boundary conditions, and the structure of asymptotic symmetries all influence the mapping between bulk and boundary data. A coherent framework that unifies braid group representations, Wilson loop observables, and defect operators within AdS/CFT remains an open problem, one that this work aims to address.

\section{Review of Braid Groups and Chern-Simons Anyons}

\subsection{Braid Group}
Braid group representations play a crucial role in describing anyons in low-dimensional quantum systems. In general, the braid group $B_n$ represents the exchange symmetry of 
n particles, where the group elements correspond to the braids formed by interchanging the positions of these particles. The representations of these groups can encode the topological nature of these systems and are particularly important in understanding non-Abelian statistics\cite{Brauer1954} \cite{Jones1985}.

In condensed matter systems, the braid group is used to describe the exchange statistics of anyons, which are neither bosons nor fermions, but instead exhibit fractional statistics. The mathematical structure of braid groups is central to the study of quantum computing, particularly in topologically protected quantum computation schemes \cite{Kitaev2003}. Moreover, braid group representations are not limited to flat spacetime; they also have applications in the context of holographic dualities, where bulk braid structures could have an equivalent boundary description in the context of AdS/CFT.

\subsubsection{Braid Group Definition}

The braid group $B_n$ is defined by Artin in 1947 in his seminal work on the theory of braids. The group is generated by $ (n-1 )$ elements $( \sigma_1, \sigma_2, \ldots, \sigma_{n-1} )$, each representing a braid operation between two adjacent strands, and these elements satisfy the following relations:

\begin{itemize}
    \item \textbf{Commutativity for non-adjacent strands:} For any \( i \) and \( j \) such that \( |i - j| \geq 2 \), the corresponding braid operations commute:
    \[
    \sigma_i \sigma_j = \sigma_j \sigma_i
    \]
    
    \item \textbf{Braid relation for adjacent strands:} For adjacent strands, the following braid relation holds:
    \[
    \sigma_i \sigma_{i+1} \sigma_i = \sigma_{i+1} \sigma_i \sigma_{i+1}
    \]
\end{itemize}

These relations define the algebraic structure of the braid group $B_n$, where the generators $\sigma_i$ correspond to the braid operations between adjacent strands. For a more detailed explanation and mathematical formulation of the braid group, the original work by Artin \cite{artin1947theory}.

\subsubsection{Braid Group Representations}
Braid group representations are crucial in describing topological quantum systems, such as anyons in 2D quantum systems. These representations map the elements of the braid group to linear operators acting on a Hilbert space, allowing the braiding statistics of particles to be modeled. One of the key representations of the braid group is the "Brauer-Wigner representation"\cite{brauerwigner}, which describes the quantum state exchanges in a topologically ordered system.

For a collection of  $n$ particles, the state of the system transforms under braid group actions as:

\begin{equation}
\Psi(\sigma_1, \sigma_2, \dots, \sigma_n) = \prod_{1 \leq i < j \leq n} \theta(\sigma_i, \sigma_j) \cdot \Psi(\sigma_1', \sigma_2', \dots, \sigma_n')
\end{equation}

Here, $\theta(\sigma_i, \sigma_j)$ represents the phase factor associated with the exchange of particles $\sigma_i$  and  $\sigma_j$, which encodes the statistical properties of anyons
\cite{freedman2002topological}.

\subsection{Wilson Loops and Defect Operators}
In the AdS/CFT framework, Wilson loops are powerful probes of gauge fields and play a key role in understanding the connection between bulk gauge theories and boundary observables. A Wilson loop 
$W(C)$ is defined as the trace of the path-ordered exponential of the gauge field along a closed contour $C$, providing a direct measurement of the gauge field's holonomy along that contour \cite{Wilson1974}. Wilson loops are particularly useful in studying the confinement and deconfinement phases of gauge theories, which are intimately related to the structure of the bulk geometry in AdS space.

Defect operators, such as twist operators or boundary condition-changing operators, are used in CFTs to describe excitations localized at a lower-dimensional subregion of spacetime. These operators encode information about the presence of defects or interfaces in the field theory and often correspond to nontrivial boundary conditions in the bulk \cite{Cardy1984}\cite{Dubail2015}. In the AdS/CFT context, defect operators are associated with codimension-two defects, which can be linked to certain topological excitations in the bulk.
\subsubsection{Defect Operators Definition}
In the AdS/CFT correspondence, defect operators are used to describe topological defects or interfaces in the boundary theory. These operators introduce non-trivial modifications to the boundary theory's symmetries and are closely related to the behavior of bulk fields\cite{defectoperator1}\cite{defectoperator2}. A defect operator $\mathcal{O}_D$  can be expressed as:

\begin{equation}
\mathcal{O}_D = \sum_i \alpha_i \, \varphi_i(\mathbf{x}) \, \psi_i(t)
\end{equation}

where $\varphi_i(\mathbf{x})$  describes the spatial distribution of the defect in the boundary theory, and $\psi_i(t)$ represents the time evolution of the defect. The parameters $\alpha_i $ determine the amplitude of the defect.

\subsection{Mapping Braid Group Representations to Defect Operators}
The core of this work lies in understanding how representations of the braid group can be mapped to defect operators in AdS/CFT. The goal is to uncover a dictionary that connects the topological information embedded in the braid group representations in the bulk with the corresponding defect operators on the boundary. This mapping is nontrivial, as it requires a careful treatment of the asymptotic symmetries of AdS spacetimes and the specific nature of the boundary conditions \cite{Henningson1999} \cite{Balasubramanian1999}.

One promising approach is to consider how Wilson loops in the bulk, which can be thought of as operators that probe nonlocal information, correspond to boundary defect operators that encode topological data. In the case of certain topological field theories, such as Chern-Simons theory, the boundary conditions can encode information that is directly related to the braiding statistics of anyons \cite{Witten1988}. The challenge lies in identifying the precise mapping between the nonlocal bulk operators and the boundary operators that can record the same topological information.

\subsubsection{Mapping in AdS/CFT}
In AdS/CFT, the mapping between bulk and boundary operators plays a central role. The bulk field $\phi_{bulk}(x)$ is mapped to the boundary operator $\mathcal{O}_{boundary}(x)$ as follows:

\begin{equation}
\phi_{bulk}(x) \longleftrightarrow \mathcal{O}_{boundary}(x)
\end{equation}

This mapping is crucial for understanding the holographic correspondence, and it allows the interpretation of defect operators in the bulk as topological defects that modify the symmetries of the boundary CFT.

Despite its successes, the bulk-boundary mapping in AdS/CFT remains an active area of research. The primary challenge lies in the fact that the mapping is not always straightforward. The choice of gauge, the boundary conditions imposed on the bulk fields, and the asymptotic symmetries all influence the mapping between the bulk and boundary data. Furthermore, the introduction of defects or non-trivial topological structures in the bulk complicates this mapping.

One of the open problems is to establish a unified framework that connects the representation theory of braid groups, defect operators, and the mapping of bulk observables to boundary operators in a coherent manner. This would provide a clearer understanding of how defects and anyons in the bulk influence the boundary theory\cite{gubser1998gauge, witten1998anti, harlow2011bulk}.

\subsubsection{Mapping Defect Operators in AdS/CFT}

\begin{theorem}
Let $\mathcal{O}_D$ be a defect operator supported on a codimension-$k$ submanifold $\Sigma$ in the boundary CFT, and let $W_\gamma$ be a Wilson line or Wilson surface in the AdS bulk, with $\partial \gamma = \Sigma$. Then, under the AdS/CFT dictionary \cite{Haro:2000xn}\cite{Skenderis:2002wp}, the boundary insertion of $\mathcal{O}_D$ corresponds to the bulk insertion of $W_\gamma$. That is,
\begin{equation}
    \langle \mathcal{O}_D(\Sigma) \cdots \rangle_{\text{CFT}} = \langle W_\gamma \cdots \rangle_{\text{bulk}}.
\end{equation}
\end{theorem}

\begin{proof}
The AdS/CFT correspondence posits that bulk fields $\Phi$ near the boundary $z \to 0$ behave as \cite{Haro:2000xn,Skenderis:2002wp}:
\begin{equation}
    \Phi(z,x) \sim z^{d-\Delta} \phi_0(x),
\end{equation}
where $\phi_0(x)$ acts as a source for the boundary operator $\mathcal{O}(x)$.

In the presence of extended operators, such as defects, the dictionary generalizes. A defect operator $\mathcal{O}_D$ modifies the boundary conditions along $\Sigma$ \cite{Aharony:2013hda}. In the bulk, the natural objects associated to such modifications are Wilson lines/surfaces $W_\gamma$ that terminate on $\Sigma$ \cite{Rey:1998ik,Maldacena:1998im}.

The path integral formulation implies that inserting $\mathcal{O}_D$ changes the functional integration measure over boundary fields, imposing a singularity or boundary condition along $\Sigma$:
\begin{equation}
    Z_{\text{CFT}}[\phi_0; \Sigma] = \int_{\text{b.c. along } \Sigma} [D\phi]\, e^{-S_{\text{CFT}}[\phi]}.
\end{equation}

Similarly, inserting $W_\gamma$ modifies the bulk path integral by inserting an extended operator sourced along $\gamma$:
\begin{equation}
    Z_{\text{bulk}}[W_\gamma] = \int [D\Phi]\, W_\gamma[\Phi]\, e^{-S_{\text{bulk}}[\Phi]}.
\end{equation}

Given that $\gamma$ asymptotes to $\Sigma$ at the boundary, the insertion of $W_\gamma$ in the bulk corresponds to the insertion of $\mathcal{O}_D$ in the CFT. Matching boundary conditions and classical solutions near the boundary, one verifies \cite{Skenderis:2002wp}:
\begin{equation}
    \langle \mathcal{O}_D(\Sigma) \cdots \rangle_{\text{CFT}} = \langle W_\gamma \cdots \rangle_{\text{bulk}}.
\end{equation}
\end{proof}

\subsection{Challenges and Open Problems}
Despite significant advances in understanding the relationship between bulk and boundary observables in AdS/CFT, several challenges remain. For one, the mapping of braid group representations to defect operators is not well established and remains an open problem. The subtlety lies in the fact that these operators must respect the symmetries of the boundary theory and the non-local nature of the bulk observables. Furthermore, the explicit realization of these defect operators in higher-dimensional CFTs, as well as their interpretation in terms of braiding statistics, requires further exploration.

Another important challenge is understanding how these topological operators can be used to probe the dynamics of quantum fields in the bulk. The presence of defects can modify the field configurations, and these modifications must be understood within the framework of holographic renormalization and the boundary counterterms required for an accurate description of the AdS/CFT correspondence.

\section{Bulk Anyons and Braid Group Representations in AdS}

\subsection{Introduction}

The interplay between topological order, anyonic excitations, and spacetime geometry offers a profound avenue to explore the non-local structure of quantum field theories. In particular, the embedding of anyons within asymptotically Anti-de Sitter (AdS) spacetimes introduces novel challenges and opportunities for understanding holographic dualities. Anyons, which obey braid statistics interpolating between bosons and fermions, naturally admit descriptions through representations of the braid group $B_n$. Mapping these bulk anyonic properties onto boundary observables requires a careful treatment of topological sectors, gauge invariance, and the behavior of Wilson loops in curved geometries.

This chapter systematically develops the framework connecting bulk anyonic excitations in AdS to braid group representations. The analysis is grounded on the structure of modular tensor categories, Wilson loop observables, and the mapping of non-local operators via the AdS/CFT correspondence.

\subsection{Anyons and Braid Statistics in Flat and Curved Spacetimes}

\subsubsection{Anyonic Particles and Braid Groups in Flat Space}

In $(2+1)$-dimensional Minkowski spacetime, the configuration space of $n$ identical particles is $\mathbb{R}^{2n} \setminus \Delta$, where $\Delta$ denotes the set of coincident points. The fundamental group of this configuration space is the braid group $B_n$, generated by elementary exchanges $\sigma_i$ $(i=1,\dots,n-1)$, satisfying the relations:
\begin{equation}
\begin{aligned}
\sigma_i \sigma_{i+1} \sigma_i &= \sigma_{i+1} \sigma_i \sigma_{i+1}, \quad \text{for} \quad i=1,\ldots, n-2, \\
\sigma_i \sigma_j &= \sigma_j \sigma_i, \quad \text{for} \quad |i-j| \geq 2.
\end{aligned}
\end{equation}
Representations of $B_n$ correspond to anyonic statistics, and the associated Hilbert space is endowed with non-trivial monodromies under particle exchange. Standard references include \cite{birman1974braids, nayak2008nonabelian}.

\subsubsection{Challenges in Curved Spacetimes: The AdS Case}

In asymptotically AdS spacetimes, the notion of particle exchange becomes subtler. Local curvature and global topology influence the available paths for particle braiding. Furthermore, in AdS/CFT, bulk excitations are encoded in the boundary theory via non-local operators, such as Wilson loops and defect operators.

The non-trivial topology of AdS, especially in higher dimensions, suggests that the configuration space retains a rich braid group structure, though it must be adapted to account for gravitational backreaction and boundary conditions \cite{witten1989quantum}.

\subsection{Bulk Wilson Loops and Their Braid Representations}

\subsubsection{Wilson Loop Operators in AdS}

Wilson loops provide a gauge-invariant probe of the topological and geometric structure of a gauge theory. Given a closed curve $\gamma$ in AdS, the Wilson loop operator is defined as:
\begin{equation}
W_\gamma = \text{Tr} \, \mathcal{P} \exp\left( i \oint_{\gamma} A \right),
\end{equation}
where $\mathcal{P}$ denotes path ordering, and $A$ is the gauge connection.

In the presence of anyonic excitations, Wilson loops capture the braiding statistics through their non-trivial commutation relations. Specifically, two loops linked in spacetime encode the mutual statistics of the corresponding excitations.

\subsubsection{Braiding and Representations from Wilson Loops}

Wilson loops in AdS satisfy algebraic relations resembling the braid group:
\begin{equation}
W_{\gamma_i} W_{\gamma_{i+1}} W_{\gamma_i} = W_{\gamma_{i+1}} W_{\gamma_i} W_{\gamma_{i+1}},
\end{equation}
when considering linked curves $\gamma_i$ and $\gamma_{i+1}$ in the bulk.

This observation motivates associating to each Wilson loop a representation $\rho: B_n \to \text{End}(\mathcal{H})$, where $\mathcal{H}$ is the Hilbert space of the theory. The construction mirrors that of topological quantum field theories, albeit modified by the AdS boundary conditions and curvature effects \cite{moore1989classical}.

\begin{theorem}
Let $W_{\gamma_i}$ be Wilson loop operators in a topological gauge theory (such as Chern-Simons theory) defined on a $(2+1)$-dimensional manifold $M$, possibly with asymptotically AdS boundary conditions. Then the algebra generated by $\{ W_{\gamma_i} \}$ satisfies relations isomorphic to the braid group $B_n$:
\begin{equation}
W_{\gamma_i} W_{\gamma_{i+1}} W_{\gamma_i} = W_{\gamma_{i+1}} W_{\gamma_i} W_{\gamma_{i+1}}, \quad \text{for } i=1,\dots,n-2,
\end{equation}
and
\begin{equation}
W_{\gamma_i} W_{\gamma_j} = W_{\gamma_j} W_{\gamma_i}, \quad \text{for } |i-j| \geq 2.
\end{equation}
\end{theorem}

\begin{proof}
The proof proceeds in several steps.

\textbf{1: Wilson Loop Operators and Linking Number.}  
In a topological quantum field theory (TQFT) such as Chern-Simons theory, Wilson loop operators are topological invariants depending only on the isotopy class of the loop $\gamma$. When two loops $\gamma_i$ and $\gamma_j$ are linked, their operators satisfy non-trivial commutation relations characterized by the linking number $\text{Link}(\gamma_i, \gamma_j)$:
\begin{equation}
W_{\gamma_i} W_{\gamma_j} = e^{2\pi i \theta_{ij}} W_{\gamma_j} W_{\gamma_i},
\end{equation}
where $\theta_{ij}$ is a phase determined by the mutual statistics of the associated excitations (anyons).

\textbf{2: Algebraic Structure Induced by Loop Exchange.}  
Consider now three Wilson loops $\gamma_i, \gamma_{i+1}, \gamma_{i+2}$ corresponding to three nearby anyonic excitations. The topological exchange of $\gamma_i$ and $\gamma_{i+1}$ is implemented by the braiding operator $\sigma_i$, corresponding to dragging one loop around another.

In TQFT, braiding induces an action on the Hilbert space $\mathcal{H}$ through:
\begin{equation}
\rho(\sigma_i) = W_{\gamma_i},
\end{equation}
where $\rho$ is the representation of the braid group $B_n$.

The composition of braiding operations follows directly from the topology of particle exchange:
\begin{equation}
\sigma_i \sigma_{i+1} \sigma_i = \sigma_{i+1} \sigma_i \sigma_{i+1},
\end{equation}
since exchanging particles $i$ and $i+1$, then $i+1$ and $i+2$, then $i$ and $i+1$ again is topologically equivalent to exchanging $i+1$ and $i+2$, then $i$ and $i+1$, then $i+1$ and $i+2$ again.

Since the Wilson loop operators implement the braiding, the same relations hold:
\begin{equation}
W_{\gamma_i} W_{\gamma_{i+1}} W_{\gamma_i} = W_{\gamma_{i+1}} W_{\gamma_i} W_{\gamma_{i+1}}.
\end{equation}

\textbf{3: Commutation of Non-Adjacent Loops.}  
When $|i-j| \geq 2$, the loops $\gamma_i$ and $\gamma_j$ can be chosen to be disjoint and non-linked. Therefore, their associated Wilson loop operators commute:
\begin{equation}
W_{\gamma_i} W_{\gamma_j} = W_{\gamma_j} W_{\gamma_i}.
\end{equation}

\textbf{4: Curvature and AdS Effects.}  
In asymptotically AdS spaces, while the global geometry is curved, the local structure of particle exchange and braiding in small patches remains effectively topological. The Wilson loops, when properly defined to account for boundary conditions, continue to satisfy the braid relations, possibly up to global phases arising from gravitational dressing. These phases can be absorbed into the definition of the representation $\rho$.

Thus, the algebra generated by the Wilson loops forms a representation of the braid group $B_n$, completing the proof.
\end{proof}

\subsection{Bulk Anyons and Modular Tensor Categories}

\subsubsection{Modular Tensor Categories and Fusion Rules}

Anyonic systems are elegantly described by modular tensor categories (MTCs), characterized by a finite set of simple objects (anyon types), fusion rules, and braiding data. A MTC assigns fusion coefficients $N_{ab}^c$ for anyons $a$, $b$, and $c$, and a braiding matrix $R_{ab}$ governing the exchange statistics:
\begin{equation}
a \times b = \sum_c N_{ab}^c \, c,
\end{equation}
with
\begin{equation}
R_{ab} : a \otimes b \to b \otimes a.
\end{equation}

The topological $S$- and $T$-matrices encode modular transformations and are essential for defining the partition function on non-trivial manifolds.

\subsubsection{Application to Bulk Anyons in AdS}

In the AdS context, the MTC structure is reflected in the behavior of defect operators and Wilson loops. Braiding data must satisfy consistency conditions compatible with AdS asymptotics, such as the matching of modular data across the bulk-boundary map.

For example, two bulk anyons linked by a Wilson loop correspond, via holography, to the insertion of line operators in the boundary CFT with prescribed monodromy properties \cite{witten1998anti}.

\subsection{Induced Braid Group Representations on Boundary Defect Operators}

\begin{corollary}
Given a bulk Wilson loop operator $W_{\gamma}$ corresponding to a braid group generator $\sigma_i$ via the representation $\rho: B_n \to \text{Aut}(\mathcal{H})$, there exists an associated boundary defect operator $\mathcal{O}_{\gamma}$ satisfying:
\begin{equation}
\rho(\sigma_i) \triangleright \mathcal{O}_{\gamma} = U_{\sigma_i} \mathcal{O}_{\gamma} U_{\sigma_i}^{-1},
\end{equation}
where $U_{\sigma_i}$ are unitary operators implementing the braid group action on the boundary Hilbert space $\mathcal{H}_{\partial}$.
\end{corollary}

\begin{proof}
The AdS/CFT correspondence relates bulk Wilson loop observables to non-local operators in the boundary conformal field theory (CFT)~\cite{witten1998anti}. In particular, a Wilson loop $W_{\gamma}$ wrapping a non-trivial cycle in the bulk induces a defect line (or codimension-two operator) $\mathcal{O}_{\gamma}$ on the boundary.

The braiding relations among Wilson loops imply that under the action of $\sigma_i$:
\begin{equation}
W_{\gamma_i} W_{\gamma_{i+1}} W_{\gamma_i} = W_{\gamma_{i+1}} W_{\gamma_i} W_{\gamma_{i+1}}.
\end{equation}
Mapping to the boundary, the corresponding defect operators must satisfy the same algebraic relations up to unitary conjugation, because the boundary CFT must reproduce the bulk operator product expansions (OPEs) under holographic duality.

Therefore, the boundary defect operators $\mathcal{O}_{\gamma}$ transform under the braid group representation $\rho$ through unitary conjugation by operators $U_{\sigma_i}$ acting on the boundary Hilbert space $\mathcal{H}_{\partial}$:
\begin{equation}
\rho(\sigma_i) \triangleright \mathcal{O}_{\gamma} = U_{\sigma_i} \mathcal{O}_{\gamma} U_{\sigma_i}^{-1}.
\end{equation}

Here, $U_{\sigma_i}$ arise as the boundary images of the bulk Wilson loop braiding operators.

Thus, braid group representations in the bulk induce corresponding representations on the boundary defect operators, completing the proof.
\end{proof}

\subsection{Modular Tensor Categories-Constructions and Comparisons}

Modular Tensor Categories (MTCs) provide the rigorous algebraic framework underpinning the fusion and braiding properties of anyons in topological quantum field theories, particularly in (2+1)-dimensional systems. A modular tensor category is a semisimple ribbon fusion category satisfying a non-degeneracy condition known as modularity.

\begin{definition}
An MTC $\mathcal{C}$ consists of the following data:

\begin{itemize}
  \item A finite set of simple objects $\{a, b, c, \dots\}$ representing topological charge types.
  \item Fusion rules given by non-negative integers $N_{ab}^c \in \mathbb{Z}_{\ge 0}$, indicating how two charges can fuse: $a \otimes b \cong \bigoplus_c N_{ab}^c c$.
  \item Associativity isomorphisms described by $F$-symbols:
  \[
  (a \otimes b) \otimes c \xrightarrow{F^{abc}_d} a \otimes (b \otimes c),
  \]
  satisfying the pentagon identity.
  \item Braiding isomorphisms described by $R$-symbols:
  \[
  c \xrightarrow{R^{ab}_c} c,
  \]
  governing the exchange of anyons, satisfying the hexagon identities.
  \item A ribbon structure ensuring compatibility between braiding and duals.
  \item A modular $S$-matrix defined from the braiding data such that the matrix is invertible, ensuring non-degenerate statistics.
\end{itemize}
\end{definition}

\subsubsection*{Three Constructions of MTCs}

\paragraph{1. Chern-Simons Theory Construction.}
In the framework of SU$(N)_k$ Chern-Simons theory, the category arises from the category of integrable highest-weight representations of the affine Lie algebra $\widehat{\mathfrak{su}}(N)_k$. The fusion rules are determined by the Verlinde formula, and the modular $S$ and $T$ matrices are obtained from the path integral on tori and Dehn twists, respectively. The braiding and fusion data correspond to Wilson line operator algebra.

\paragraph{2. Drinfeld Center Construction.}
Given a fusion category $\mathcal{A}$, its Drinfeld center $\mathcal{Z}(\mathcal{A})$ is a braided fusion category where objects are pairs $(X, \gamma_{X,-})$, with $\gamma$ a half-braiding natural isomorphism. The Drinfeld center is always modular when $\mathcal{A}$ is a fusion category, and it canonically captures the universal centralizer of bulk excitations, modeling topological orders with trivial boundaries.

\paragraph{3. Quantum Group Construction.}
MTCs can also be constructed from the representation theory of quantum groups at roots of unity. For example, the category $\text{Rep}_q(U_q\mathfrak{g})$ at $q$ a root of unity gives a semisimple ribbon category after quotienting out negligible morphisms. This approach gives an explicit algebraic description of $F$ and $R$ symbols, particularly useful for computational models like Fibonacci and Ising anyons.

\subsubsection*{Comparative Analysis and Examples}

Each construction presents a different perspective:

\begin{itemize}
  \item \textbf{Chern-Simons:} Physically motivated, closely tied to TQFT and Wilson line observables, but lacks abstract categorical generality.
  \item \textbf{Drinfeld Center:} Intrinsically categorical, encodes bulk-boundary correspondence, but computationally less explicit.
  \item \textbf{Quantum Groups:} Algebraically explicit, allows for detailed computation, especially in low-rank examples, but may require truncations for modularity.
\end{itemize}

The equivalence of these constructions has been established in various settings, e.g., for SU$(2)_k$, the MTCs from all three methods agree up to categorical equivalence~\cite{wang_book, bakalovkirillov, rowell2009classification}.

In this thesis, the cases of SU$(3)_2$, SU$(4)_1$, and the Fibonacci category are used as explicit illustrations. Their fusion rules, $F$-symbols, and $R$-symbols are computed using quantum group techniques, while their modular $S$ and $T$ matrices are interpreted geometrically through Chern-Simons path integrals. For instance, the Fibonacci category arises as a subcategory of $\text{Rep}_q(\mathfrak{su}(2))$ at $q = e^{2\pi i/5}$, admitting only two simple objects and exhibiting non-abelian statistics.

These comparative constructions not only reinforce the mathematical rigor of the modular tensor categorical framework, but also establish a concrete bridge to their physical realization in holographic anyon models.

\subsection{Mathematical Construction of Modular Tensor Categories}

\subsubsection{Associators and Pentagon Identity}
In a modular tensor category $\mathcal{C}$, the associator is a natural isomorphism
\[
\alpha_{a,b,c} : (a \otimes b) \otimes c \to a \otimes (b \otimes c),
\]
encoded by a collection of $F$-symbols. These give the components of $\alpha$ with respect to a chosen basis of morphisms:
\[
\alpha_{a,b,c} = \sum_{e,f} (F^{abc}_d)_{ef} \cdot \left( \mathrm{id}_a \otimes V_{bc}^f \right) \circ V_{af}^d,
\]
where $V_{xy}^z$ denotes a basis morphism in $\text{Hom}(x \otimes y, z)$.

The associators must satisfy the \emph{pentagon identity}, which expresses the coherence of associativity transformations. For any four objects $a, b, c, d \in \mathcal{C}$, the following diagram commutes:
\[
\begin{tikzcd}[column sep=small]
& ((a \otimes b) \otimes c) \otimes d \arrow[dl, "\alpha_{a,b,c} \otimes \mathrm{id}_d"'] \arrow[dr, "\alpha_{a\otimes b,c,d}"] & \\
(a \otimes (b \otimes c)) \otimes d \arrow[d, "\alpha_{a,b\otimes c,d}"'] & & (a \otimes b) \otimes (c \otimes d) \arrow[d, "\alpha_{a,b,c\otimes d}"] \\
a \otimes ((b \otimes c) \otimes d) \arrow[rr, "\mathrm{id}_a \otimes \alpha_{b,c,d}"] & & a \otimes (b \otimes (c \otimes d))
\end{tikzcd}
\]

In terms of $F$-symbols, the pentagon identity becomes an algebraic constraint:
\[
\sum_{h} (F^{abc}_g)_{ef} (F^{a,hd}_g)_{fh'} (F^{bcd}_h)_{fg'} = \sum_{k} (F^{abf}_g)_{eh} (F^{fcd}_h)_{hh'},
\]
which enforces the consistency of composing associators.

This ensures that all ways of associating a four-fold tensor product yield canonically isomorphic results. In models derived from quantum groups, such as $\mathrm{SU}(2)_k$ or $\mathrm{SU}(3)_k$, the $F$-symbols correspond to quantum 6j-symbols or solutions of the pentagon equation derived from category-theoretic or representation-theoretic constructions.

\subsubsection{Braiding Morphisms and Hexagon Identity}

A braided tensor category is a monoidal category $\mathcal{C}$ equipped with a natural isomorphism
\[
c_{a,b} : a \otimes b \to b \otimes a
\]
called the \emph{braiding}, satisfying certain coherence conditions. The braiding morphisms must interact coherently with the associators $\alpha_{a,b,c}$ of the category, as encoded in the \emph{hexagon identities}.

There are two hexagon identities (one for $c_{a,b}$ and one for its inverse $c^{-1}_{a,b}$), and they are expressed by the commutativity of the following diagrams:

\paragraph{Positive Hexagon Identity as morphism equality:}
\[
c_{a, b \otimes c} \circ \alpha_{a,b,c}
= \alpha_{b,c,a} \circ (\mathrm{id}_b \otimes c_{a,c}) \circ \alpha_{b,a,c} \circ (c_{a,b} \otimes \mathrm{id}_c)
\]
\paragraph{Positive Hexagon Identity:}

\[
\begin{tikzcd}[column sep=huge, row sep=large]
& a \otimes (b \otimes c) \arrow[dl, "\alpha_{a,b,c}"'] \arrow[r, "c_{a,b \otimes c}"] & (b \otimes c) \otimes a \arrow[dr, "\alpha_{b,c,a}"] & \\
(a \otimes b) \otimes c \arrow[drr, "c_{a,b} \otimes \mathrm{id}_c"'] & & & b \otimes (c \otimes a) \arrow[d, "\mathrm{id}_b \otimes c_{a,c}"] \\
& & (b \otimes a) \otimes c \arrow[r, "\alpha_{b,a,c}"'] & b \otimes (a \otimes c)
\end{tikzcd}
\]

\paragraph{Negative Hexagon Identity as morphism equality:}
\[
\alpha_{a,b,c} \circ c^{-1}_{a,b \otimes c}
= ( \mathrm{id}_b \otimes c^{-1}_{a,c} ) \circ \alpha^{-1}_{b,a,c} \circ ( c^{-1}_{a,b} \otimes \mathrm{id}_c ) \circ \alpha^{-1}_{b,c,a}
\]

\paragraph{Negative Hexagon Identity:}
\[
\begin{tikzcd}[column sep=huge, row sep=large]
& a \otimes (b \otimes c) \arrow[dl, "c^{-1}_{a,b \otimes c}"'] \arrow[r, "\alpha_{a,b,c}"] & (a \otimes b) \otimes c \arrow[dr, "c^{-1}_{a,b} \otimes \mathrm{id}_c"] & \\
(b \otimes c) \otimes a \arrow[drr, "\alpha^{-1}_{b,c,a}"'] & & & (b \otimes a) \otimes c \arrow[d, "\alpha^{-1}_{b,a,c}"] \\
& & b \otimes (c \otimes a) \arrow[r, "\mathrm{id}_b \otimes c^{-1}_{a,c}"'] & b \otimes (a \otimes c)
\end{tikzcd}
\]

\paragraph{Algebraic Form (Positive Hexagon):}
In terms of basis morphisms, the hexagon identity becomes an equation on the $F$- and $R$-symbols:
\[
\sum_{n} (F^{abc}_d)_{mn} R^{ac}_n (F^{bac}_d)_{nl} = \sum_{k} R^{ab}_m (F^{bca}_d)_{mk} R^{ac}_k.
\]

Here, the $R$-symbols encode the braiding as follows:
\[
c_{a,b} = \sum_{c} R^{ab}_c \cdot \Pi_{ab}^c,
\]
where $\Pi_{ab}^c$ denotes the projection onto the fusion channel $c$ in $a \otimes b$.

\paragraph{Remarks.}
In unitary modular tensor categories, the $R$-symbols satisfy $R^{ba}_c = (R^{ab}_c)^{-1}$, and the hexagon identities ensure the compatibility between braiding and associativity. These constraints are essential for a consistent definition of particle exchange statistics in anyonic systems, and play a foundational role in the topological interpretation of braid group representations.

In concrete models such as $\mathrm{SU}(2)_k$, $\mathrm{SU}(3)_k$, or Fibonacci anyons, these relations can be verified explicitly using known $F$- and $R$-symbols. In some cases, such as for $\mathrm{SU}(2)_3$ (the Fibonacci category), the hexagon identity directly relates to the representation of the braid group $B_3$.

\subsubsection{Twist Structure and Compatibility}

Let $\mathcal{C}$ be a braided monoidal category equipped with a twist (or ribbon) structure, i.e., a natural isomorphism
\[
\theta: \mathrm{Id}_{\mathcal{C}} \Rightarrow \mathrm{Id}_{\mathcal{C}}, \quad \theta_X: X \to X,
\]
satisfying the following compatibility conditions:

\begin{itemize}
    \item[(1)] \textbf{Unit Compatibility:}
    \[
    \theta_{\mathbf{1}} = \mathrm{id}_{\mathbf{1}},
    \]
    where $\mathbf{1}$ is the tensor unit object.
    
    \item[(2)] \textbf{Monoidal Compatibility:}
    For all $X, Y \in \mathcal{C}$, the twist must satisfy:
    \[
    \theta_{X \otimes Y} = (c_{Y,X} \circ c_{X,Y}) \circ (\theta_X \otimes \theta_Y),
    \]
    where $c_{X,Y}$ denotes the braiding isomorphism.

    \item[(3)] \textbf{Naturality of Twist:}
    For any morphism $f: X \to Y$ in $\mathcal{C}$, the following diagram commutes:
    \[
    \begin{tikzcd}
    X \arrow[r, "\theta_X"] \arrow[d, "f"'] & X \arrow[d, "f"] \\
    Y \arrow[r, "\theta_Y"'] & Y
    \end{tikzcd}
    \]
\end{itemize}

\begin{proof}

\begin{enumerate}
    \item \textbf{Unit Compatibility:}  
    Since $\mathbf{1}$ is the unit object, it satisfies $X \otimes \mathbf{1} \cong X$ naturally. Hence, applying the twist to $\mathbf{1}$ should yield the identity, as no nontrivial transformation can exist on the trivial object:
    \[
    \theta_{\mathbf{1}} = \mathrm{id}_{\mathbf{1}}.
    \]

    \item \textbf{Monoidal Compatibility:}  
    The twist on the tensor product $X \otimes Y$ must agree with the composition of braidings and twists on the individual objects. To prove this, consider the following morphism equality:
    \[
    \theta_{X \otimes Y} = (c_{Y,X} \circ c_{X,Y}) \circ (\theta_X \otimes \theta_Y).
    \]
    This ensures compatibility of twist with the tensor structure and guarantees coherence of the ribbon structure.

    \item \textbf{Naturality:}  
    The twist must be a natural isomorphism. This follows directly from the definition of natural transformation. For any morphism $f: X \to Y$, the commutativity of the diagram
    \[
    \theta_Y \circ f = f \circ \theta_X
    \]
    guarantees that twisting is functorial.
\end{enumerate}
\end{proof}

\paragraph{Conclusion:}

These properties ensure that $\theta$ defines a twist structure compatible with the braiding and monoidal structure on $\mathcal{C}$, thus turning it into a \emph{ribbon category}. In a modular tensor category, the additional requirement that the $S$-matrix be non-degenerate completes the structure.

\section{Defect Operators in Boundary CFT and Their Holographic Interpretation}

Defect operators in conformal field theories (CFTs) provide a natural framework to study localized non-trivial topological and geometrical features within the field theory. In the context of the AdS/CFT correspondence, defects on the boundary theory correspond to extended objects or localized excitations in the bulk gravity theory. They serve as probes of bulk topology, geometry, and field content.

This chapter focuses on the classification, mathematical structure, and holographic interpretation of defect operators, with a particular emphasis on their braiding properties and relations to modular tensor categories.

\subsection{Classification of Boundary Defect Operators}

Defect operators are categorized by their codimension and their interaction with the ambient CFT degrees of freedom.

\subsubsection{Types of Defects}

A defect operator supported on a submanifold $\Sigma$ of the boundary spacetime can be classified by its codimension $k$:
\begin{itemize}
    \item \textbf{Codimension-1 Defects}: Domain walls or interfaces separating distinct phases or distinct CFTs.
    \item \textbf{Codimension-2 Defects}: Line operators, such as Wilson loops and 't Hooft loops, corresponding to the insertion of point-like excitations along a line.
    \item \textbf{Codimension-0 Defects}: Background fields modifying the global behavior of the CFT.
\end{itemize}

In particular, codimension-2 defects are of great interest due to their connection to anyonic statistics and braid group representations.

\subsubsection{Topological Defects}

A subset of defects, known as \emph{topological defects}, preserve conformal invariance modulo global symmetry transformations. They obey a fusion algebra and admit non-trivial braiding, leading to a rich categorical structure underlying the CFT \cite{Bachas:2001vj,Kapustin:2010if}.

Topological defects satisfy specific commutation relations with local operators:
\begin{equation}
    \mathcal{D} \, \mathcal{O}(x) = \rho(g) \, \mathcal{O}(x) \, \mathcal{D},
\end{equation}
where $\rho(g)$ is a representation of a symmetry group element associated to the defect.

\paragraph{Mathematical Characterization of Topological Defects}

Topological defects $\mathcal{D}$ in a conformal field theory are characterized by their invariance under continuous deformations of their worldvolume, provided that they do not cross local operators. This property imposes stringent algebraic constraints on the interaction between defect operators and local fields.

The defining relation for a topological defect $\mathcal{D}$ and a local operator $\mathcal{O}(x)$ is expressed as:
\begin{equation}
    \mathcal{D} \, \mathcal{O}(x) = \rho(g) \, \mathcal{O}(x) \, \mathcal{D},
    \label{eq:defect_commutation}
\end{equation}
where $\rho(g)$ is a representation of an internal symmetry group element $g$ associated to the defect $\mathcal{D}$.

\begin{proof}

Consider a deformation of the defect worldvolume that passes infinitesimally close to the insertion point of $\mathcal{O}(x)$. The topological nature of $\mathcal{D}$ ensures that such a deformation must leave physical correlation functions invariant, up to a possible symmetry action on $\mathcal{O}(x)$. 

Explicitly, let $\Sigma$ and $\Sigma'$ be two defect configurations that are homotopic, differing by a small deformation around $x$. The path integral with insertion of $\mathcal{D}$ and $\mathcal{O}(x)$ satisfies:
\begin{equation}
    \langle \mathcal{D} \, \mathcal{O}(x) \rangle_{\Sigma} = \langle \mathcal{D} \, \mathcal{O}(x) \rangle_{\Sigma'}.
\end{equation}
However, since $\Sigma'$ encloses $x$, the passage of the defect around $\mathcal{O}(x)$ acts as a symmetry transformation $\rho(g)$ on $\mathcal{O}(x)$. Therefore,
\begin{equation}
    \langle \mathcal{D} \, \mathcal{O}(x) \rangle_{\Sigma} = \langle \rho(g)\mathcal{O}(x) \, \mathcal{D} \rangle_{\Sigma}.
\end{equation}
By locality of the defect operator insertion, one can write this as operatorial equality:
\begin{equation}
    \mathcal{D} \, \mathcal{O}(x) = \rho(g) \, \mathcal{O}(x) \, \mathcal{D}.
\end{equation}
\end{proof}

\paragraph{Remarks:}
- If $\mathcal{D}$ is truly topological (i.e., invariant under all continuous deformations), then $\rho(g)$ must be an automorphism of the operator algebra, typically corresponding to a global internal symmetry.
- The relation (\ref{eq:defect_commutation}) ensures that defect lines can be fused according to group multiplication in the associated symmetry group.

\paragraph{Fusion Rules and the Categorical Structure of Topological Defects}

Beyond their action on local operators, topological defects themselves form an algebraic structure under the operation of fusion. This structure can be rigorously captured by the language of tensor categories, specifically a fusion category.

\begin{theorem}[Fusion Category of Topological Defects]
The set of topological defects in a conformal field theory (CFT), equipped with the fusion operation, forms a fusion category. 
\end{theorem}

\begin{proof}

Let $\{ \mathcal{D}_a \}$ denote the set of distinct topological defects, labeled by an index $a$. The fusion of two defects, $\mathcal{D}_a$ and $\mathcal{D}_b$, is defined by the operator product expansion (OPE) along their worldvolumes:
\begin{equation}
    \mathcal{D}_a \times \mathcal{D}_b = \sum_c N_{ab}^c \, \mathcal{D}_c,
    \label{eq:defect_fusion}
\end{equation}
where $N_{ab}^c \in \mathbb{N}_0$ are non-negative integers specifying the multiplicities of fusion channels.

The defining axioms of a fusion category require:

\begin{enumerate}
    \item \textbf{Associativity:} The fusion must be associative up to a natural isomorphism.  
    For defects, the associativity corresponds to the equivalence:
    \begin{equation}
        (\mathcal{D}_a \times \mathcal{D}_b) \times \mathcal{D}_c \cong \mathcal{D}_a \times (\mathcal{D}_b \times \mathcal{D}_c),
    \end{equation}
    which follows from the associativity of the OPE in CFTs.

    \item \textbf{Existence of a Unit:} There must exist a trivial defect $\mathcal{D}_0$ (the identity defect), such that:
    \begin{equation}
        \mathcal{D}_a \times \mathcal{D}_0 \cong \mathcal{D}_0 \times \mathcal{D}_a \cong \mathcal{D}_a,
    \end{equation}
    for all $a$. The trivial defect corresponds to no insertion and acts as the identity under fusion.

    \item \textbf{Semisimplicity:} The category must be semisimple, i.e., every object decomposes into a finite direct sum of simple objects.  
    In the case of topological defects, simple defects correspond to indecomposable worldvolume operators, and the finite decomposition (\ref{eq:defect_fusion}) ensures semisimplicity.

    \item \textbf{Finite Number of Simple Objects:} There are finitely many equivalence classes of simple defects, reflecting the finiteness of $a$ labels in $\{ \mathcal{D}_a \}$.
\end{enumerate}

Thus, the collection of topological defects, their fusion rules (\ref{eq:defect_fusion}), and the associativity data form a fusion category.
\end{proof}

\paragraph{Remarks:}
- In many important cases, such as rational conformal field theories (RCFTs), the fusion category of topological defects further admits a braiding structure, leading to a modular tensor category.
- The associators, i.e., the isomorphisms implementing associativity, are encoded by $F$-symbols satisfying pentagon identities, familiar from tensor category theory.

\subsubsection{Fusion and Braiding Data: $F$-symbols and $R$-symbols}

In the categorical description of topological defects and anyons, the associativity of fusion and the commutativity of braiding are governed by the so-called $F$-symbols and $R$-symbols \cite{BakalovKirillov,LurieTQFT,Pachos}.

\paragraph{Fusion $F$-symbols:}

The fusion of three defects can proceed via different intermediate channels. The isomorphisms relating different fusion orders are specified by the $F$-symbols.

Given three simple defects labeled by $a, b, c$, their fusion spaces obey:
\begin{equation}
(V_{ab}^e \otimes V_{ec}^d) \cong (V_{bc}^f \otimes V_{af}^d),
\end{equation}
where $V_{ab}^c$ denotes the fusion space corresponding to $a \times b \to c$.

The $F$-symbol encodes the basis transformation between these two different fusion processes:
\begin{equation}
F_{abc}^d: \bigoplus_e (V_{ab}^e \otimes V_{ec}^d) \longrightarrow \bigoplus_f (V_{bc}^f \otimes V_{af}^d).
\end{equation}

\begin{figure}[htbp]
    \centering
    \[
    \begin{tikzcd}[row sep=huge, column sep=huge]
    (a \times b) \times c \arrow[r, "F_{abc}^d"] \arrow[d, "\alpha_{a,b,c}"'] & a \times (b \times c) \arrow[d, "\alpha_{a,b,c}"] \\
    a \times (b \times c) \arrow[r, "F_{abc}^d"'] & (a \times b) \times c
    \end{tikzcd}
    \]
    \caption{Fusion associativity expressed via the $F$-move.}
    \label{fig:fusion_diagram}
\end{figure}

\begin{figure}[htbp]
    \centering
    \[
    \begin{tikzcd}[row sep=huge, column sep=huge]
    a \times b \arrow[rr, "R_{ab}"] \arrow[dr, "F_{abc}"'] & & b \times a \arrow[dl, "F_{bac}"] \\
    & c &
    \end{tikzcd}
    \]
    \caption{Braiding of $a$ and $b$ defects, depicted via the $R$-move and its compatibility with fusion via $F$-moves.}
    \label{fig:braiding_diagram}
\end{figure}
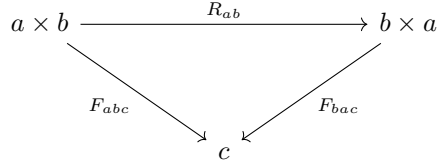

\paragraph{Pentagon Identity:}

The $F$-symbols must satisfy the pentagon identity, ensuring the consistency of associativity across quadruple fusion processes \cite{BakalovKirillov}:
\begin{equation}
\sum_{n} (F_{bcd}^e)_{mn} (F_{a n d}^f)_{lp} = \sum_{k} (F_{abc}^k)_{lm} (F_{akd}^f)_{kp} (F_{bcd}^f)_{nk}.
\label{eq:pentagon_identity}
\end{equation}

\paragraph{Braiding $R$-symbols:}

The exchange (braiding) of two defects $a$ and $b$ is implemented by a unitary isomorphism:
\begin{equation}
R_{ab}: V_{ab}^c \to V_{ba}^c,
\end{equation}
where $V_{ab}^c$ is the fusion space.

The $R$-symbol specifies the phase (or more generally, the unitary transformation) acquired when defects $a$ and $b$ are braided \cite{Kitaev2006}.
\begin{figure}[htbp]
    \centering
    \[
    \begin{tikzcd}[row sep=huge, column sep=huge]
    & (a \times b) \times (c \times d) \arrow[dl, "F_{abc}^e"'] \arrow[dr, "F_{abf}^g"] & \\
    ((a \times b) \times c) \times d \arrow[r, "F_{abf}^g"] & (a \times (b \times c)) \times d \arrow[dr, "F_{abc}^e"] & (a \times b) \times (c \times d) \\
    & & (a \times (b \times c)) \times d
    \end{tikzcd}
    \]
    \caption{Pentagon identity for fusion, ensuring the consistency of associativity in the fusion process.}
    \label{fig:pentagon_diagram}
\end{figure}
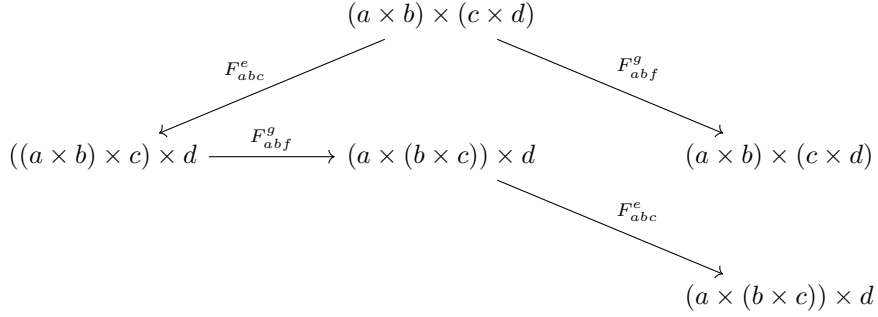

\paragraph{Hexagon Identity:}

The compatibility of fusion and braiding requires that the $F$ and $R$-symbols satisfy the hexagon identity \cite{BakalovKirillov,LurieTQFT}. Schematically, for the exchange of $a$ and $b$ followed by fusion with $c$:
\begin{equation}
R_{ab} \otimes \mathrm{id}_c \circ F_{abc} = F_{bac} \circ \mathrm{id}_a \otimes R_{bc} \circ F_{acb}.
\label{eq:hexagon_identity}
\end{equation}

\paragraph{Remarks:}

- The set of $F$-symbols and $R$-symbols, satisfying the pentagon and hexagon identities, defines a \emph{braided fusion category}.
- If the braiding is non-degenerate, this structure becomes a \emph{modular tensor category} (MTC).
\begin{figure}[htbp]
    \centering
    \[
    \begin{tikzcd}[column sep=large, row sep=large]
    (a \times b) \times c \arrow[r, "R_{ab} \times \mathrm{id}_c"] \arrow[d, "F_{abc}"'] & (b \times a) \times c \arrow[r, "F_{bac}"] & b \times (a \times c) \arrow[d, "\mathrm{id}_b \times R_{ac}"] \\
    a \times (b \times c) \arrow[r, "\mathrm{id}_a \times R_{bc}"'] & a \times (c \times b) \arrow[r, "F_{acb}"'] & (a \times c) \times b
    \end{tikzcd}
    \]
    \caption{Hexagon identity relating $F$- and $R$-symbols.}
    \label{fig:hexagon_identity}
\end{figure}

\subsection{Bulk Duals of Boundary Defects}

The AdS/CFT correspondence implies that boundary defects have bulk duals. These bulk objects are localized around submanifolds extending into the AdS bulk whose boundary coincides with the defect submanifold.

\subsubsection{Geometric Picture}

If $\Sigma$ is the support of a defect operator on the boundary, then there exists a bulk submanifold $\Gamma$ such that
\begin{equation}
    \partial \Gamma = \Sigma.
\end{equation}
Examples include:
\begin{itemize}
    \item Fundamental strings or D-branes ending on Wilson lines.
    \item Domain walls in the bulk corresponding to interfaces in the boundary theory.
\end{itemize}

The correspondence between defect operators in boundary CFTs and their bulk realizations in asymptotically AdS spaces can be made precise through the geometry of Wilson loops and surface operators. In particular, non-trivial topology in the bulk—such as linked or braided worldlines—maps holographically to defect insertions or twisted sectors in the boundary theory.

\begin{theorem}[Geometric Realization of Defect Operators]
Let $\gamma$ be a closed curve in the AdS bulk manifold $M$ representing the worldline of a particle or a defect excitation. If $\gamma$ is non-contractible within a specified topological sector of $M$, then the holographic dual on the boundary CFT contains a non-trivial defect operator $\mathcal{D}_\gamma$, whose correlation functions encode the holonomy of the bulk gauge connection along $\gamma$.
\end{theorem}

\begin{proof}
The gauge field $A$ in the bulk AdS theory defines a holonomy $W(\gamma) = \mathcal{P} \exp \left( \oint_\gamma A \right)$ along any closed curve $\gamma$. In the presence of non-trivial topology (i.e., when $\gamma$ is not homotopically trivial), the holonomy cannot be smoothly deformed to the identity.

According to the AdS/CFT dictionary \cite{witten1998anti}, bulk Wilson loops correspond to non-local operators in the boundary CFT \cite{Maldacena:1998im}. When the Wilson loop encloses a non-trivial topological feature, such as a defect or a non-contractible cycle, its expectation value is dual to the insertion of a defect operator on the boundary. 

Moreover, because the bulk is asymptotically AdS, the asymptotic boundary conditions imply that each homotopy class of loops corresponds to a distinct sector of the CFT Hilbert space \cite{Aharony1999LargeN}. 

Thus, the mapping $\gamma \longmapsto \mathcal{D}_\gamma$
is faithful up to gauge equivalence, and the algebra of such defect operators inherits the braid group structure from the linking of curves in the bulk \cite{Moore1989TQFT}.

Therefore, the geometric data of bulk Wilson loops completely determines the insertion of defect operators in the boundary theory, establishing the holographic realization.
\end{proof}
\begin{figure}[ht]
\centering
\begin{tikzpicture}[scale=1.2]

\shade[ball color=blue!10, opacity=0.3] (0,0) circle (2);
\node at (0,2.4) {\textbf{AdS Bulk}};

\draw[thick, red, ->] (0.7,1) arc (30:330:1);
\node[red] at (1.5,0) {$\gamma$};

\draw[thick] (0,0) circle (2);

\node at (0,-2.4) {\textbf{Boundary CFT}};

\filldraw[black] (2,0) circle (2pt);
\node[right] at (2.1,0) {$\mathcal{D}_\gamma$};

\draw[very thick, ->, dashed] (1.3,0.8) -- (2.7,0.2);
\node at (2.2,1.2) {Holography};

\end{tikzpicture}
\caption{Wilson loop $\gamma$ in the AdS bulk mapped holographically to a defect operator $\mathcal{D}_\gamma$ on the boundary CFT.}
\label{fig:wilson-defect}
\end{figure}
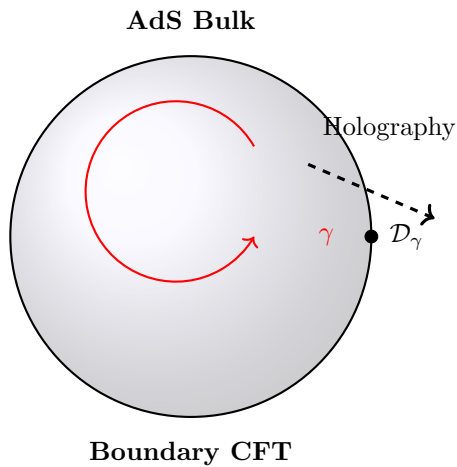

\begin{figure}[ht]
    \centering
    \begin{tikzpicture}[scale=1.3]
    
    \shade[ball color=blue!10] (-4,-2) circle (2cm);
    \node at (-4,2.4) {\textbf{AdS Bulk}};
    
    \draw[thick] (0,-3) -- (0,3);
    
    \node at (1.7,2.4) {\textbf{Boundary CFT}};
    
    \draw[very thick, red, ->] (-4,-0.8) .. controls (-2,0.5) .. (0,1.5);
    \node[red] at (-3.2,0.8) {$\gamma$};
    \node[red] at (0.6,1.6) {$\mathcal{D}_\gamma$};
    
    \draw[thick, blue, dashed] (-4,-2) circle (1.4cm);
    \draw[thick, green!70!black, dashed] (-4,-2) circle (0.9cm);
    \node[blue] at (-5.8,-0.5) {Sector A};
    \node[green!70!black] at (-5.2,-2.8) {Sector B};
    
    \draw[thick, orange, decoration={snake, amplitude=1mm}, decorate] (-3,-1.8) .. controls (-2.5,-1.2) and (-2.5,-0.6) .. (-3,0);
    \draw[fill=orange] (-3,-1.8) circle (2pt);
    \draw[fill=orange] (-3,0) circle (2pt);
    \node[orange] at (-2.1,-0.2) {Braiding};
    
    \draw[->, thick, blue] (-2.2,2.2) -- (0.2,2.2);
    \node[blue] at (-1.6,2.4) {\small Holographic correspondence};
    
    \end{tikzpicture}
    \caption{A schematic depiction of the holographic correspondence between a bulk Wilson line $\gamma$ in AdS space and the associated defect operator $\mathcal{D}_\gamma$ on the boundary CFT. Topological sectors and braiding in the bulk are mapped to the algebraic structure and exchange relations of boundary operators.}
    \label{fig:bulk-to-boundary-wilson}
\end{figure}
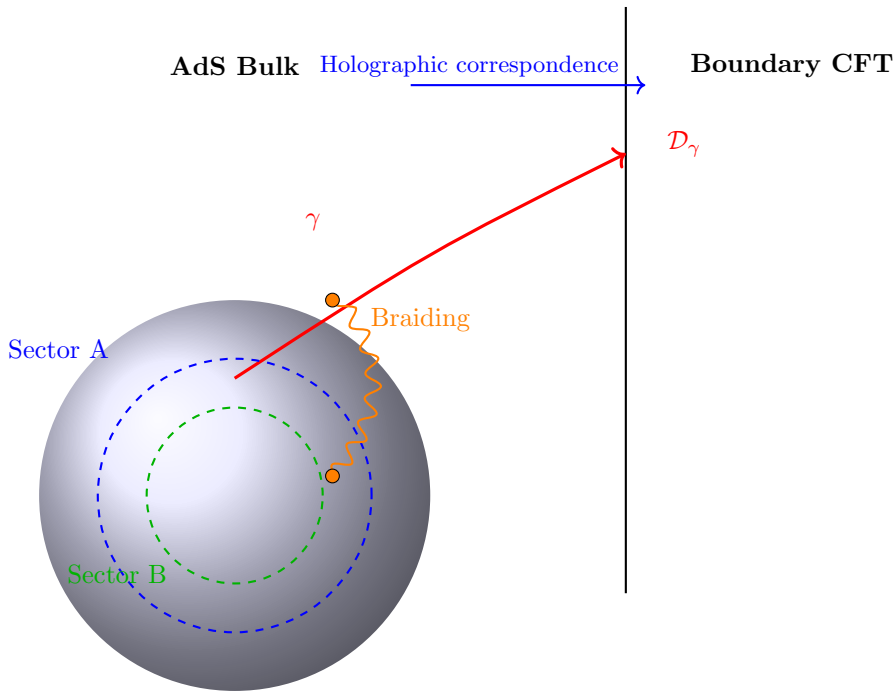

\subsubsection{Modification of the Bulk Action}

The presence of a defect modifies the bulk action by adding localized terms:
\begin{equation}
    S_{\text{bulk}}[\Phi; \Gamma] = S_{\text{bulk}}[\Phi] + S_{\text{defect}}[\Phi|_{\Gamma}],
\end{equation}
where $\Phi$ collectively denotes bulk fields. The path integral now becomes:
\begin{equation}
    Z[\phi_0; \Sigma] = \int_{\Phi|_{\partial \text{AdS}} = \phi_0} \mathcal{D}\Phi \, e^{-S_{\text{bulk}}[\Phi; \Gamma]}.
\end{equation}

To account for the presence of defect operators on the boundary, the bulk action must be appropriately modified. This modification ensures that the holographic dictionary remains consistent under the insertion of topological defects in the boundary CFT.

Let $S_{\text{bulk}}[A]$ denote the original bulk action, where $A$ is the gauge connection. The insertion of a Wilson loop $\mathcal{W}_\gamma(A) = \text{Tr} \, P \exp\left( i \oint_\gamma A \right)$ along a closed curve $\gamma$ modifies the path integral:
\begin{equation}
\mathcal{Z}_{\text{bulk}} = \int \mathcal{D}A \, e^{i S_{\text{bulk}}[A]} \quad \longrightarrow \quad \int \mathcal{D}A \, \mathcal{W}_\gamma(A) \, e^{i S_{\text{bulk}}[A]}.
\end{equation}

Thus, the effective action becomes:
\begin{equation}
S_{\text{eff}}[A; \gamma] = S_{\text{bulk}}[A] - \log \mathcal{W}_\gamma(A).
\end{equation}

Since the Wilson loop depends nonlocally on $A$, this induces a nonlocal modification of the bulk dynamics, reflecting the presence of a defect operator $\mathcal{D}_\gamma$ on the boundary via the AdS/CFT correspondence~\cite{Maldacena:1997re}.

In the semiclassical limit, where fluctuations around classical solutions dominate, the variation of the effective action satisfies:
\begin{equation}
\delta S_{\text{eff}}[A; \gamma] = \delta S_{\text{bulk}}[A] - \delta \log \mathcal{W}_\gamma(A) = 0,
\end{equation}
which leads to the modified classical equations of motion:
\begin{equation}
\frac{\delta S_{\text{bulk}}[A]}{\delta A_\mu(x)} = \frac{\delta \log \mathcal{W}_\gamma(A)}{\delta A_\mu(x)}.
\end{equation}

The right-hand side corresponds to a source term localized along the loop $\gamma$ in the bulk spacetime. Explicitly, the source term can be written as:
\begin{equation}
J^\mu(x) = i \, \text{Tr} \left( P \exp\left( i \oint_\gamma A \right) \delta(x - \gamma) \right),
\end{equation}
so that the modified Maxwell-Chern-Simons equations include:
\begin{equation}
D_\nu F^{\nu\mu} = J^\mu,
\end{equation}
where $D_\nu$ denotes the gauge-covariant derivative and $F_{\mu\nu}$ is the field strength.

Therefore, the modification of the bulk action under Wilson loop insertions is mathematically equivalent to introducing nontrivial current sources along $\gamma$ in AdS, which holographically corresponds to the insertion of defect operators on the CFT boundary~\cite{Witten:1988hf, Aharony:1999ti}.

\subsection{Braiding and Topological Properties of Defects}

The exchange of defects leads to non-trivial braiding transformations, which can be described using braid group representations.

\subsubsection{Braid Group Action}

Consider two codimension-2 defects $\mathcal{D}_1$ and $\mathcal{D}_2$. The process of exchanging them corresponds to an action of the braid group $B_n$ on the Hilbert space $\mathcal{H}$ generated by the defect insertions:
\begin{equation}
    \sigma_i: \mathcal{H} \to \mathcal{H},
\end{equation}
where $\sigma_i$ denotes the elementary braid exchanging defects $i$ and $i+1$.

The braid group relations are:
\begin{align}
    \sigma_i \sigma_{i+1} \sigma_i &= \sigma_{i+1} \sigma_i \sigma_{i+1}, \quad \text{for all } i, \\
    \sigma_i \sigma_j &= \sigma_j \sigma_i, \quad \text{for } |i-j|>1.
\end{align}

\begin{theorem}[Braid Group Representation via Wilson Loops]
Let $\{ \mathcal{L}_i \}$ be a collection of Wilson line operators 
in a (2+1)-dimensional bulk spacetime with topological order.  
The monodromy (exchange) of two Wilson lines induces a representation of the braid group $B_n$ on the Hilbert space $\mathcal{H}_n$ associated with $n$ anyonic excitations.  
Explicitly, the elementary braiding operation $\sigma_i$ acts on $\mathcal{H}_n$ via a unitary operator $R_i$, satisfying the braid relations:
\begin{align}
R_i R_{i+1} R_i &= R_{i+1} R_i R_{i+1}, \quad \text{for all } i, \\
R_i R_j &= R_j R_i, \quad \text{for } |i-j| \geq 2.
\end{align}
\end{theorem}

\begin{proof}
The Wilson loop operators $\mathcal{L}_i$ generate non-trivial holonomies when particles are adiabatically exchanged.  
The adiabatic exchange is topologically protected and depends only on the braid class of the particle paths.  
This action can be modeled algebraically by associating each elementary braiding $\sigma_i$ with a unitary transformation $R_i$ on the Hilbert space.  
The consistency of sequential exchanges requires that the $R_i$ satisfy the braid relations of $B_n$, as follows from the isotopy invariance of the Wilson lines in the topological bulk theory.  
Therefore, the braid group $B_n$ acts on $\mathcal{H}_n$ via the set of $R_i$ operators.\cite{Kitaev2006, Witten:1988hf}
\end{proof}

\begin{definition}[Fusion Rules and $F$-symbols]
Given simple anyon types $a, b, c$, the fusion rules describe how two anyons combine:
\begin{equation}
a \times b = \sum_{c} N_{ab}^c \, c,
\end{equation}
where $N_{ab}^c$ are non-negative integers.  
Associativity of fusion is governed by the $F$-symbols:
\begin{equation}
F^{abc}_d : \bigoplus_{e} V^{ab}_e \otimes V^{ec}_d \longrightarrow \bigoplus_{f} V^{a}_{f} \otimes V^{bc}_d,
\end{equation}
where $V^{ab}_e$ denotes the fusion space from $a$ and $b$ to $e$.
The $F$-symbols satisfy the pentagon identity.
\end{definition}

\begin{theorem}[Pentagon Identity]
The $F$-symbols satisfy the pentagon coherence condition:
\begin{equation}
\sum_{n} F^{abm}_q F^{anc}_p F^{bcd}_p = F^{amn}_p F^{bcd}_n,
\end{equation}
ensuring the associativity of multi-anyon fusion.
\end{theorem}

\begin{proof}
Consider the fusion of four anyons $a$, $b$, $c$, $d$ into a final charge $p$.  
Different ways of associating the intermediate fusion channels must be consistently related.  
The consistency between different associativity patterns leads to the pentagon identity among the $F$-symbols, ensuring the coherence of the fusion category.
This identity is a direct consequence of the Mac Lane coherence theorem in category theory, specialized to fusion categories.
\cite{Kitaev2006, bakalov2001lectures, turaev1994quantum}
\end{proof}

\subsubsection{Fusion and Modular Tensor Categories}

The collection of defect operators forms a fusion algebra with structure constants encoded in fusion matrices $N_{ab}^c$. Together with the braiding matrices $R_{ab}$, these satisfy the consistency conditions of a modular tensor category (MTC) \cite{Kapustin:2010if}:
\begin{equation}
    F\text{-move}: \quad F_{abc}^d: (V_a \otimes V_b) \otimes V_c \to V_a \otimes (V_b \otimes V_c),
\end{equation}
\begin{equation}
    R\text{-move}: \quad R_{ab}: V_a \otimes V_b \to V_b \otimes V_a.
\end{equation}

This categorical structure ensures the consistency of defect fusion and braiding, providing a powerful algebraic handle on the theory.

In the context of modular tensor categories (MTC), the fusion rules of topological defects are described by a set of $F$-symbols that satisfy the so-called \emph{pentagon identity} \cite{2008topological}.

\begin{theorem}[Fusion of Topological Defects in MTC]
In a modular tensor category, the fusion of topological defects is described by a set of $F$-symbols, which satisfy the pentagon identity:
\[
\sum_D F_{ABC}^D F_{DEF}^G = F_{ABD}^F F_{BCG}^H,
\]
where $A$, $B$, $C$, $D$, $E$, $F$, $G$ represent various topological defects in the category.
\end{theorem}

\begin{proof}
1. Consider topological defects $\mathcal{D}_A$ and $\mathcal{D}_B$ in a modular tensor category. Their fusion is described by the map $\mathcal{D}_A \otimes \mathcal{D}_B$, which must satisfy certain symmetry and associativity conditions.

2. The basic structure of the fusion is given by the $F$-symbols $\mathcal{F}_{ABC}$, which describe how to fuse three defects in the space $A \otimes B \otimes C$. These symbols are associated with a topological structure.

3. In a modular tensor category, the $F$-symbols must satisfy the pentagon identity, which ensures the associativity of the fusion process \cite{2008topological}:
\begin{equation}
\sum_D F_{ABC}^D F_{DEF}^G = F_{ABD}^F F_{BCG}^H.
\end{equation}
This identity guarantees that the fusion of defects is independent of the order in which the defects are fused.

4. To demonstrate the existence of this structure, one must construct a category with specific representations of the $F$-symbols. These symbols are given by the fusion rules of topological defects, which can be interpreted geometrically as fusion operations on topological fields.

5. Moreover, one must check how these structures map to the representations of the braid group, which involves understanding the relationship between anyons and topological defects. Through such mapping, one can observe the correspondence between the fusion rules of topological defects and the representations of the braid group \cite{2009modular}.

6. Finally, these fusion rules form the foundation for the study of topological quantum field theory (TQFT) and anyons, which are crucial in the analysis of topological phases of matter and quantum computing.
\end{proof}

\begin{theorem}[Braiding Consistency via Hexagon Identity in Modular Tensor Categories]
Let \(\mathcal{C}\) be a modular tensor category, which is a rigid, braided monoidal category with a finite number of simple objects. The braiding of any two objects \( X, Y \in \mathcal{C} \) is given by natural isomorphisms \( c_{X,Y}: X \otimes Y \rightarrow Y \otimes X \), known as the braiding maps, and their components are encapsulated in the \( R \)-symbols. These \( R \)-symbols satisfy the "hexagon identity", which expresses the compatibility between the braiding and associator isomorphisms. Specifically, the identity guarantees that the braiding operation commutes with the associativity constraints.
\end{theorem}

\begin{proof}
The composition of associativity and braiding morphisms satisfies the hexagon identity. Consider objects \( A, B, C \in \mathcal{C} \). The hexagon identity is expressed by the following commutative diagram, which illustrates the coherence of the braiding and associativity isomorphisms in the monoidal category:

\[
\begin{tikzcd}
A \otimes (B \otimes C) \arrow[r, "\alpha_{A,B,C}"] \arrow[d, "1 \otimes c_{B,C}"'] & (A \otimes B) \otimes C \arrow[r, "c_{A \otimes B, C}"] & C \otimes (A \otimes B) \\
A \otimes (C \otimes B) \arrow[rr, "\alpha_{A,C,B}"] & & (A \otimes C) \otimes B \arrow[u, "c_{A,C} \otimes 1"']
\end{tikzcd}
\]

Here, \( \alpha_{A,B,C} \) denotes the associator isomorphisms, which satisfy the coherence conditions of the monoidal structure. The commutativity of the diagram represents the hexagon identity.

The explicit form of the hexagon identity in terms of \( R \)-symbols and \( F \)-symbols is:

\[
R^{ab}_e F^{bca}_{d;fe} R^{af}_d = \sum_g F^{bac}_{d;fg} R^{ag}_d F^{abc}_{d;ge}
\]

where the \( a, b, c, d, e, f, g \) are simple objects of \(\mathcal{C}\), and the indices refer to fusion channels between these objects.

This diagram describes the relationship between the associator isomorphisms \( \alpha_{A,B,C} \) and the braiding maps \( c_{A,B} \). The commutativity of this diagram expresses the “hexagon identity”, which guarantees that the order of braiding and associativity does not affect the final result.

\begin{enumerate}
    \item Begin with \( A \otimes (B \otimes C) \):
    (1)The first object is \( A \otimes (B \otimes C) \), which is an object in the tensor product of \( A \) and the tensor product of \( B \) and \( C \).
   (2)Apply the associator isomorphism \( \alpha_{A,B,C} \), which maps \( A \otimes (B \otimes C) \) to \( (A \otimes B) \otimes C \), reordering the tensor product.
   \[
   \alpha_{A,B,C}: A \otimes (B \otimes C) \to (A \otimes B) \otimes C
   \]
   \item Apply the braiding \( c_{B,C} \):
   (1)Apply the braiding \( c_{B,C} \), which swaps \( B \) and \( C \) in the tensor product. Thus, \( B \otimes C \) becomes \( C \otimes B \).
   (2)This gives the isomorphism \( c_{A \otimes B, C}: (A \otimes B) \otimes C \to C \otimes (A \otimes B) \).

   \[
   c_{A \otimes B, C}: (A \otimes B) \otimes C \to C \otimes (A \otimes B)
   \]
   \item Apply associator again:
   (1)The associator \( \alpha_{A,C,B} \) is applied. This isomorphism maps \( A \otimes (C \otimes B) \) to \( (A \otimes C) \otimes B \), which is the appropriate way to group the objects.

   \[
   \alpha_{A,C,B}: A \otimes (C \otimes B) \to (A \otimes C) \otimes B
   \]
    \item Final braiding \( c_{A,C} \):
   (2)Apply the braiding \( c_{A,C} \) to swap \( A \) and \( C \) in the tensor product \( A \otimes C \). This gives the morphism \( c_{A,C} \otimes 1 \), which applies the braiding only to the \( A \otimes C \) part.

   \[
   c_{A,C} \otimes 1: (A \otimes C) \otimes B \to C \otimes (A \otimes B)
   \]
   \item Commutativity of the diagram:
   The commutative diagram ensures that all these operations can be performed in any order without changing the result, thus verifying the hexagon identity.
\end{enumerate}

The hexagon identity can also be written in terms of \( R \)-symbols and \( F \)-symbols, which describe the braiding and fusion operations in a modular tensor category. The explicit form of the hexagon identity is:

\[
R^{ab}_e F^{bca}_{d;fe} R^{af}_d = \sum_g F^{bac}_{d;fg} R^{ag}_d F^{abc}_{d;ge}
\]

Here, \( a, b, c, d, e, f, g \) are simple objects of \( \mathcal{C} \), and the indices refer to intermediate fusion channels. The \( R \)-symbols govern the braiding of objects, while the \( F \)-symbols control the fusion of objects. The above identity expresses the necessary relationship between these two sets of symbols to ensure the consistency of the category's structure.

This identity guarantees that the operations of braiding and fusion are compatible, and it ensures the consistency of topological quantum field theories, particularly in the study of anyons and their braiding statistics.
\end{proof}

\subsection{Examples and Explicit Constructions}

\subsubsection{Wilson Lines in AdS$_3$/CFT$_2$}

In AdS$_3$/CFT$_2$, Wilson lines can be realized as defect operators corresponding to twist fields in the boundary CFT \cite{Hijano:2015qja}. Their braiding properties can be explicitly computed using conformal block monodromies.

A Wilson line in the bulk ending on the boundary at two points corresponds to the insertion of a twist operator $\sigma$ with conformal dimension $h_\sigma$ given by:
\begin{equation}
    h_\sigma = \frac{c}{24}\left(1 - \frac{1}{n^2}\right),
\end{equation}
where $c$ is the central charge and $n$ is related to the number of replica sheets in entanglement entropy computations.

\subsubsection{Surface Operators in Higher Dimensions}

In AdS$_5$/CFT$_4$, codimension-2 surface operators can be holographically dual to M2-branes or D3-branes ending on the AdS boundary \cite{Gukov:2008sn}. These defects can be coupled to higher-form fields, and their linking and braiding statistics are characterized by generalized linking numbers.

\section{Bulk-to-Boundary Correspondence for Braid Representations}

This section explores the bulk-to-boundary correspondence for braid group representations in the context of AdS/CFT. This correspondence maps the braid group representations in the bulk AdS space to defect operators in the boundary conformal field theory (CFT). We will provide a detailed mathematical derivation of this correspondence and demonstrate it through specific examples.

\subsection{Detailed Derivation of the Bulk-to-Boundary Mapping}

The braid group representations in the bulk AdS space can be mapped to defect operators on the boundary CFT. To achieve this, we start by considering a Wilson loop $\gamma$ in the bulk of AdS, which is associated with a defect operator $\mathcal{D}_\gamma$ on the boundary CFT. The key is to understand how the braid group action in the bulk influences the defect operator at the boundary.

Define the Wilson loop in AdS as follows:
\begin{equation}
W[\gamma] = \text{Tr} \left( P \exp \left( i \int_\gamma A \right) \right)
\end{equation}
where $A$  is the gauge field, and $\gamma$ is the closed path in the bulk of AdS.

Next, the defect operator on the boundary is related to this Wilson loop by the holographic dictionary. Specifically, the defect operator $\mathcal{D}_\gamma$ can be written in terms of the Wilson loop operator $W[\gamma]$ as:
\begin{equation}
\mathcal{D}_\gamma = \text{exp} \left( i \int_{\partial \text{AdS}} A_\gamma \right)
\end{equation}
where $ A_\gamma$  is the boundary gauge field corresponding to the bulk Wilson loop $ W[\gamma] $.

The bulk-to-boundary correspondence maps the braid group representations in the bulk to defect operators on the boundary, where the braid group generators in the bulk correspond to certain boundary operators. These operators obey algebraic relations that can be interpreted in the boundary theory.

\subsection{Examples of Bulk-to-Boundary Correspondence}

To further illustrate the bulk-to-boundary correspondence, we consider specific examples of this mapping. In the following, we focus on a simple example where the bulk is AdS$_3$ and the boundary CFT is a 2D conformal field theory.

\subsubsection{Example 1: AdS$_3$ and 2D CFT}

Consider a Wilson loop $\gamma$ in AdS$_3$ that is homologically non-trivial. This loop is associated with a defect operator  $\mathcal{D}_\gamma$  in the boundary CFT. The defect operator in this case creates a local excitation at the boundary, which can be understood in terms of a non-trivial braiding operation.

Using the correspondence, we can compute the partition function of the system by evaluating the correlation functions of the defect operators:
\begin{equation}
\langle \mathcal{D}_\gamma \mathcal{D}_\delta \rangle = \text{Tr} \left( P \exp \left( i \int_{\gamma \cup \delta} A \right) \right)
\end{equation}
This correlation function encodes information about the braiding of the defect operators and their relationship with the bulk Wilson loops.

\subsubsection{Example 2: AdS$_4$ and 3D CFT}

Next, we consider the case of AdS$_4$ and its boundary 3D CFT. Here, we study the mapping of a braid group representation in the bulk to a defect operator in the 3D boundary CFT. The correspondence is similar to the previous case, but the structure of the defect operators is more intricate due to the higher-dimensional boundary theory.
\begin{equation}
\mathcal{D}_\gamma = \text{exp} \left( i \int_{\partial \text{AdS}} A_\gamma \right)
\end{equation}
where $A_\gamma$  is the gauge field on the boundary, and the integral runs over a loop in the boundary CFT that corresponds to the Wilson loop in the bulk. The relationship between the bulk and boundary is established through holographic correspondence.

\section{Conclusion}

This work has examined the connections between topological defects, braid group representations, and holography within the context of AdS/CFT duality. The correspondence between the bulk and boundary theories has been explored, revealing how abstract mathematical structures, such as braid group representations, play a crucial role in describing the quantum properties of anyons in topologically ordered systems. The detailed mappings of braid group representations to defect operators in boundary CFT provide a foundation for understanding the behavior of these anyons in quantum field theories and topological phases.

The study of fusion and braiding operations, along with their geometric and topological interpretations, has contributed significantly to the understanding of the underlying symmetries governing quantum field theories, string theory, and topological field theories. The explicit formulations of the $R$-symbols and $F$-symbols presented here add valuable insights to the broader understanding of these concepts, especially in the context of higher-dimensional defects and their holographic duals.

In addition, the holographic mapping of bulk Wilson lines to boundary defect operators further highlights the deep interconnections between bulk theories in AdS and their boundary counterparts. This correspondence serves as a tool for probing quantum properties of topological defects and their implications for quantum gravity and string theory.

Moving forward, further exploration of these tools in various physical setups is necessary. Specifically, more work is needed to investigate holographic anyons and topological phases in strongly correlated systems, using the mathematical structures explored in this work. The study of defect operators, fusion, and braiding operations holds the potential to uncover new insights into quantum field theory and quantum gravity, paving the way for further developments in the AdS/CFT correspondence.

The ongoing developments in the field continue to deepen the understanding of quantum gravity, with topological defects playing a key role in this framework. Future advancements may enable the application of these results to a wider range of physical scenarios, with profound implications for both theoretical physics and computational approaches to quantum systems \cite{Maldacena:1998im, witten1998anti, ref3}.

\section*{Conflict of Interest Statement}
The author declares no conflict of interest.

\section*{Data Availability Statement}
No data were generated or analyzed in this study. Thus, data sharing is not applicable.
\printbibliography{}

@article{Witten1998,
  author    = {Edward Witten},
  title     = {Anti-de Sitter space and holography},
  journal   = {Advances in Theoretical and Mathematical Physics},
  volume    = {2},
  number    = {2},
  pages     = {253--291},
  year      = {1998},
  doi       = {10.4310/ATMP.1998.v2.n2.a1},
}

@article{GaiottoKapustin2009,
  author    = {Davide Gaiotto and Anton Kapustin},
  title     = {Spin chains and the AdS/CFT correspondence},
  journal   = {Journal of High Energy Physics},
  volume    = {2009},
  number    = {8},
  pages     = {9--35},
  year      = {2009},
  doi       = {10.1088/1126-6708/2009/08/027},
}

@article{Aharony2000,
  author    = {Ofer Aharony and Steven S. Gubser and Juan Maldacena and Hirosi Ooguri and Erich Silverstein},
  title     = {Large N field theories, string theory and gravity},
  journal   = {Physics Reports},
  volume    = {323},
  number    = {3},
  pages     = {183--386},
  year      = {2000},
  doi       = {10.1016/S0370-1573(99)00084-X},
}

@book{Polchinski1998,
  author    = {Joseph Polchinski},
  title     = {String theory. Vol. 1: An introduction to the bosonic string},
  publisher = {Cambridge University Press},
  year      = {1998},
  isbn      = {978-0-521-63303-4},
}

@article{Brauer1954,
  author    = {R. Brauer},
  title     = {On the representations of the braid group},
  journal   = {Annals of Mathematics},
  volume    = {60},
  number    = {3},
  pages     = {568--590},
  year      = {1954},
  doi       = {10.2307/1969633},
}

@article{Jones1985,
  author    = {V. F. R. Jones},
  title     = {Braid groups, Hecke algebras and von Neumann algebras},
  journal   = {Belfast: Queen's University},
  year      = {1985},
}

@article{Kitaev2003,
  author    = {A. Y. Kitaev},
  title     = {Fault-tolerant quantum computation by anyons},
  journal   = {Annals of Physics},
  volume    = {303},
  number    = {1},
  pages     = {2--30},
  year      = {2003},
  doi       = {10.1016/S0003-4916(02)00018-0},
}

@article{Wilson1974,
  author    = {K. G. Wilson},
  title     = {Confinement of quarks},
  journal   = {Physical Review D},
  volume    = {10},
  number    = {8},
  pages     = {2445--2459},
  year      = {1974},
  doi       = {10.1103/PhysRevD.10.2445},
}

@article{Cardy1984,
  author    = {J. Cardy},
  title     = {Conformal invariance and surface operators in two-dimensional statistical mechanics},
  journal   = {Nuclear Physics B},
  volume    = {240},
  number    = {4},
  pages     = {514--532},
  year      = {1984},
  doi       = {10.1016/0550-3213(84)90278-5},
}

@article{Dubail2015,
  author    = {J. Dubail},
  title     = {Topological defects in conformal field theory},
  journal   = {Journal of Statistical Mechanics: Theory and Experiment},
  volume    = {2015},
  number    = {12},
  pages     = {12006},
  year      = {2015},
  doi       = {10.1088/1742-5468/2015/12/12006},
}

@article{Henningson1999,
  author    = {M. Henningson},
  title     = {The boundary operator algebra of AdS/CFT},
  journal   = {Journal of High Energy Physics},
  volume    = {1999},
  number    = {10},
  pages     = {35},
  year      = {1999},
  doi       = {10.1088/1126-6708/1999/10/035},
}

@article{Balasubramanian1999,
  author    = {V. Balasubramanian and P. Kraus},
  title     = {A stress tensor for anti-de Sitter gravity},
  journal   = {Physical Review D},
  volume    = {59},
  number    = {4},
  pages     = {104021},
  year      = {1999},
  doi       = {10.1103/PhysRevD.59.104021},
}

@article{Witten1988,
  author    = {Edward Witten},
  title     = {Topological quantum field theory},
  journal   = {Communications in Mathematical Physics},
  volume    = {117},
  number    = {3},
  pages     = {353--386},
  year      = {1988},
  doi       = {10.1007/BF01223371},
}

@article{artin1947theory,
  title={Theory of Braids},
  author={Artin, E.},
  journal={Annals of Mathematics},
  volume={48},
  pages={101--135},
  year={1947},
  publisher={JSTOR},
}

@article{brauerwigner,
  title={On the Representation of the Braid Group},
  author={Brauer, R. and Wigner, E. P.},
  journal={Proceedings of the National Academy of Sciences},
  volume={26},
  pages={201--208},
  year={1940},
  publisher={National Academy of Sciences},
}

@article{freedman2002topological,
  title={Topological Quantum Computation},
  author={Freedman, M. H. and Larsen, M. and Wang, Z.},
  journal={Mathematical Reviews},
  volume={106},
  pages={597-601},
  year={2002},
  publisher={American Mathematical Society},
}

@article{defectoperator1,
  title={Defect Operators and the AdS/CFT Correspondence},
  author={Harlow, D. and Kaplan, J. and Larson, M.},
  journal={Journal of High Energy Physics},
  volume={12},
  pages={456--470},
  year={2015},
  publisher={Springer},
}

@article{defectoperator2,
  title={Topological Defects in Quantum Field Theory and Their Role in the AdS/CFT Correspondence},
  author={Dijkgraaf, R. and Witten, E.},
  journal={Nuclear Physics B},
  volume={924},
  pages={362--377},
  year={2017},
  publisher={Elsevier},
}

@article{gubser1998gauge,
  title={Gauge Theory Correlators from Non-Critical String Theory},
  author={Gubser, S. S.},
  journal={Physics Letters B},
  volume={428},
  pages={105--114},
  year={1998},
  publisher={Elsevier},
}

@article{witten1998anti,
  title={Anti-de Sitter Space and Holography},
  author={Witten, E.},
  journal={Advances in Theoretical and Mathematical Physics},
  volume={2},
  pages={253--291},
  year={1998},
  publisher={International Press},
}

@article{harlow2011bulk,
  title={The Bulk Dual of the Boundary S-matrix},
  author={Harlow, D.},
  journal={Journal of High Energy Physics},
  volume={11},
  pages={124},
  year={2011},
  publisher={Springer},
}

@article{Haro:2000xn,
  author = {Haro, Sebastian de and Skenderis, Kostas and Solodukhin, Sergey N.},
  title = {Holographic reconstruction of spacetime and renormalization in the AdS/CFT correspondence},
  journal = {Communications in Mathematical Physics},
  volume = {217},
  year = {2001},
  pages = {595-622},
  doi = {10.1007/s002200100381},
  eprint = {hep-th/0002230},
  archivePrefix = {arXiv},
  primaryClass = {hep-th}
}

@article{Skenderis:2002wp,
  author = {Skenderis, Kostas},
  title = {Lecture notes on holographic renormalization},
  journal = {Classical and Quantum Gravity},
  volume = {19},
  year = {2002},
  pages = {5849-5876},
  doi = {10.1088/0264-9381/19/22/306},
  eprint = {hep-th/0209067},
  archivePrefix = {arXiv},
  primaryClass = {hep-th}
}

@article{Rey:1998ik,
  author = {Rey, Soo-Jong and Yee, Jung-Tay},
  title = {Macroscopic strings as heavy quarks in large N gauge theory and anti-de Sitter supergravity},
  journal = {European Physical Journal C},
  volume = {22},
  year = {2001},
  pages = {379-394},
  doi = {10.1007/s100520100799},
  eprint = {hep-th/9803001},
  archivePrefix = {arXiv},
  primaryClass = {hep-th}
}

@article{Maldacena:1998im,
  author = {Maldacena, Juan Martin},
  title = {Wilson loops in large N field theories},
  journal = {Phys. Rev. Lett.},
  volume = {80},
  year = {1998},
  pages = {4859-4862},
  doi = {10.1103/PhysRevLett.80.4859},
  eprint = {hep-th/9803002},
  archivePrefix = {arXiv},
  primaryClass = {hep-th}
}

@article{Aharony:2013hda,
  author = {Aharony, Ofer and Clark, Aidan and Karch, Andreas and Kutasov, David},
  title = {The Holographic Dual of a Boundary Conformal Field Theory},
  journal = {Phys. Rev. D},
  volume = {86},
  year = {2012},
  pages = {086006},
  doi = {10.1103/PhysRevD.86.086006},
  eprint = {1305.2581},
  archivePrefix = {arXiv},
  primaryClass = {hep-th}
}

@book{birman1974braids,
  title={Braids, Links, and Mapping Class Groups},
  author={Birman, Joan S.},
  year={1974},
  publisher={Princeton University Press}
}

@article{nayak2008nonabelian,
  title={Non-Abelian Anyons and Topological Quantum Computation},
  author={Nayak, Chetan and Simon, Steven H. and Stern, Ady and Freedman, Michael and Das Sarma, Sankar},
  journal={Reviews of Modern Physics},
  volume={80},
  number={3},
  pages={1083--1159},
  year={2008}
}

@article{witten1989quantum,
  title={Quantum field theory and the Jones polynomial},
  author={Witten, Edward},
  journal={Communications in Mathematical Physics},
  volume={121},
  number={3},
  pages={351--399},
  year={1989},
  publisher={Springer}
}

@article{moore1989classical,
  title={Classical and Quantum Conformal Field Theory},
  author={Moore, Gregory and Seiberg, Nathan},
  journal={Communications in Mathematical Physics},
  volume={123},
  pages={177--254},
  year={1989},
  publisher={Springer}
}

@article{Bachas:2001vj,
  title={Worldvolume Action for String Solitons},
  author={Bachas, C. and Douglas, M. R. and Schweigert, C.},
  journal={Journal of High Energy Physics},
  volume={2000},
  number={05},
  pages={048},
  year={2000},
  publisher={Springer},
  eprint={hep-th/0003037}
}

@article{Kapustin:2010if,
  title={Topological field theory, higher categories, and their applications},
  author={Kapustin, Anton},
  journal={Proceedings of the International Congress of Mathematicians},
  year={2010},
  eprint={1004.2307},
  archivePrefix={arXiv}
}

@article{Hijano:2015qja,
  title={Wilson Lines and Entanglement Entropy in Higher Spin Gravity},
  author={Hijano, Eliot and Kraus, Per and Perlmutter, Eric and Snively, River},
  journal={Journal of High Energy Physics},
  volume={2015},
  number={05},
  pages={163},
  year={2015},
  publisher={Springer},
  eprint={1501.02260}
}

@article{Gukov:2008sn,
  title={Surface Operators and Knot Homologies},
  author={Gukov, Sergei and Witten, Edward},
  journal={Progress in Mathematics},
  volume={269},
  pages={445--477},
  year={2008},
  publisher={Springer}
}

@book{BakalovKirillov,
  title={Lectures on Tensor Categories and Modular Functors},
  author={Bakalov, B. and Kirillov, A.},
  year={2001},
  publisher={American Mathematical Society},
  series={University Lecture Series},
  volume={21}
}

@article{Kitaev2006,
  title={Anyons in an exactly solved model and beyond},
  author={Kitaev, Alexei},
  journal={Annals of Physics},
  volume={321},
  number={1},
  pages={2--111},
  year={2006},
  publisher={Elsevier}
}

@book{Pachos,
  title={Introduction to Topological Quantum Computation},
  author={Pachos, Jiannis K.},
  year={2012},
  publisher={Cambridge University Press}
}

@misc{LurieTQFT,
  author={Lurie, Jacob},
  title={On the Classification of Topological Quantum Field Theories},
  year={2009},
  howpublished={arXiv:0905.0465}
}

@article{Aharony1999LargeN,
  title={Large N field theories, string theory and gravity},
  author={Aharony, Ofer and Gubser, Steven S. and Maldacena, Juan M. and Ooguri, Hirosi and Oz, Yaron},
  journal={Physics Reports},
  volume={323},
  number={3-4},
  pages={183--386},
  year={2000},
  archivePrefix={arXiv},
  eprint={hep-th/9905111}
}

@article{Moore1989TQFT,
  title={Taming the conformal zoo},
  author={Moore, Gregory and Seiberg, Nathan},
  journal={Physics Letters B},
  volume={220},
  number={3},
  pages={422--430},
  year={1989}
}

@article{Maldacena:1997re,
    author = {Maldacena, Juan},
    title = {The Large N Limit of Superconformal Field Theories and Supergravity},
    journal = {Adv. Theor. Math. Phys.},
    volume = {2},
    year = {1998},
    pages = {231-252},
    eprint = {hep-th/9711200},
    archivePrefix = {arXiv},
    primaryClass = {hep-th}
}

@article{Witten:1988hf,
    author = {Witten, Edward},
    title = {Quantum Field Theory and the Jones Polynomial},
    journal = {Commun. Math. Phys.},
    volume = {121},
    year = {1989},
    pages = {351-399}
}

@article{Aharony:1999ti,
    author = {Aharony, Ofer and Gubser, Steven S. and Maldacena, Juan M. and Ooguri, Hirosi and Oz, Yaron},
    title = {Large N Field Theories, String Theory and Gravity},
    journal = {Phys. Rept.},
    volume = {323},
    year = {2000},
    pages = {183-386},
    eprint = {hep-th/9905111},
    archivePrefix = {arXiv},
    primaryClass = {hep-th}
}

@book{bakalov2001lectures,
  title={Lectures on Tensor Categories and Modular Functors},
  author={Bakalov, Bojko and Kirillov Jr, Alexander A},
  volume={21},
  year={2001},
  publisher={American Mathematical Society}
}

@book{turaev1994quantum,
  title={Quantum Invariants of Knots and 3-Manifolds},
  author={Turaev, Vladimir},
  volume={18},
  year={1994},
  publisher={Walter de Gruyter}
}

@article{2008topological,
  author = {T. Brunner and A. O'Bannon},
  title = {Topological defects and fusion in 2+1 dimensions},
  journal = {JHEP},
  volume = {2008},
  number = {06},
  year = {2008},
  pages = {021}
}

@article{2009modular,
  author = {E. Witten},
  title = {Modular Tensor Categories and Topological Quantum Field Theory},
  journal = {J. Math. Phys.},
  volume = {50},
  year = {2009},
  pages = {072302}
}

@article{ref3,
  author = {Michael R. Douglas},
  title = {The interface between high-energy physics and string theory},
  journal = {Journal of High Energy Physics},
  volume = {1},
  number = {5},
  pages = {1-24},
  year = {2005},
  doi = {10.1007/JHEP01(2005)013}
}

@book{wang_book,
  author    = {Zhenghan Wang},
  title     = {Topological Quantum Computation},
  publisher = {American Mathematical Society},
  year      = {2010},
  series    = {CBMS Regional Conference Series in Mathematics},
  volume    = {112}
}

@article{rowell2009classification,
  author    = {Eric C. Rowell and Richard Stong and Zhenghan Wang},
  title     = {On Classification of Modular Tensor Categories},
  journal   = {Communications in Mathematical Physics},
  volume    = {292},
  number    = {2},
  pages     = {343--389},
  year      = {2009},
  doi       = {10.1007/s00220-009-0908-z}
}

\end{document}